\documentclass{ws-procs975x65}

\def\beq{\begin{equation}}
\def\eeq{\end{equation}}

\begin{document}

\title{Super-Massive Black Hole mass estimation from bright flares}
\author{Vladim\'{\i}r Karas,\footnote{E-mail: vladimir.karas@cuni.cz} Michal Bursa, Michal Dov\v{c}iak}
\address{Astronomical Institute, Czech Academy of Sciences,\\
Bo\v{c}n\'{\i} II 1401, CZ-14100 Prague, Czech Republic}

\author{Andreas Eckart,\footnote{Also at Max-Planck-Institut f\"ur Radioastronomie, Bonn, Germany} Monika Valencia-S, Munawwar Khanduwala, \& Michal Zaja\v{c}ek$^\dag$\footnote{New address: Center for Theoretical Physics, Polish Academy of Sciences, Warsaw, Poland}} 
\address{I. Physikalisches Institut, Universit\"at zu K\"oln,\\ Z\"ulpicher Str. 77, D-50937 Cologne, Germany}

\begin{abstract}
Super-Massive Black Holes reside in galactic nuclei, where they exhibit episodic bright flares due to accretion events. Taking into account relativistic effects, namely, the boosting and lensing of X-ray flares, we further examine the possibility to constraint the mass of the SMBH from the predicted profiles of the observed light curves. To this end, we have studied four bright flares from Sagittarius A*, which exhibit an asymmetric shape consistent with a combination of two intrinsically separate peaks that occur with a specific time delay with respect to each other. We thus proposed that an interplay of relativistic effects could be responsible for the shape of the observed light curves and we tested the reliability of the method (Karssen et al. 2017, Mon. Not. R. Astron. Soc. 472, 4422; arXiv:1709.09896).
\end{abstract}
\bodymatter

\section{Introduction}
Astrophysical black holes influence their cosmic environs by gravitational and electromagnetic effects. They are  described by
a small number of free parameters, namely, the mass and angular momentum.\citep{Mis-Tho-Whe:1973} All evidence as well as theoretical arguments 
point to the fact that the effects of electric charge must be very small (at least in the steady-state limit), whereas the magnetic monopole and additional even more exotic 
parameters are thought to be negligible or they vanish completely to zero.\citep{Kov:2010,Zaj:2018a,Zaj:2018b} Several independent methods have been  
developed to measure the black hole mass as a primary parameter defining the black hole action on luminous matter in their
neigborhood.\citep{Kor:1992,Kor:2013,Cze:2010} A new method of mass determination from optical polarization of broad emission lines
has been proposed to constrain SMBH mass in active galactic nuclei.\citep{Afa:2015,Sav:2018} In X-rays, also the possibility of measuring
the black hole mass from the energy shift of narrow spectral lines has been discussed widely.\citep{Dov:2004c,Pech:2005,Kar:2010,Soch:2011}
We describe a recently developed method that is based on the assumption that a fast orbital motion of radiating X-ray flares
can be detected in time variability of accreting black holes.\citep{Kar:2017,Eck:2018} 

\section{Bright flares from a source orbiting near a black hole}
The time-scale of matter orbiting along $r=\mbox{const}$
circular trajectory around Kerr black hole of mass $M$
can be written
\begin{equation}
T_{\rm{}orb}(r;a)\simeq310(r^\frac{3}{2}+a)M_7\ {\rm sec}.
\end{equation}
Hereafter we use dimensionless geometric units, where length is expressed in terms
of gravitational radius, $R_g{\equiv}G{M}/c^2\,{\simeq}\,1.5\times10^{12}M_7$~cm
and the spin parameter $-1\leq a\leq1$ (positive values for the prograde rotation
with the black hole). The
dimension-less angular momentum $a$ adopts values in the range
$-1\leq{a}\leq1$. Positive values correspond to co-rotating motion, 
while negative values describe counter-rotation (many papers assume that
the accretion disc co-rotates, although such an assumption may not be
necessarily true). Circular orbits of free particles are possible above
the innermost stable circular orbit (ISCO a.k.a. marginally stable orbit).\cite{Bar:1972}
Gravitation governs the orbital motion near the
horizon, and so the apparent variability associated with the motion can be
scaled with the black hole mass. 

The gravitational field is described by the metric of Kerr black hole 
\cite{Mis-Tho-Whe:1973}
\begin{equation}
ds^2=-\frac{\Delta}{\Sigma}\Big(dt-a\sin^2\theta\;d\phi\Big)^2
+\frac{\Sigma}{\Delta}\;dr^2+\Sigma\;d\theta^2
+\frac{\sin^2\theta}{\Sigma}\Big[a\;dt-\big(r^2+a^2\big)\;d\phi\Big]^2
\label{eq:metric}
\end{equation}
in Boyer-Lindquist (spheroidal) coordinates $t$, $r$, $\theta$, $\phi$. 
The metric functions $\Delta(r)$
and $\Sigma(r,\theta)$ are known in an explicit form. The event horizon occurs at 
the roots of equation $\Delta(r)=0$; the 
outer solution is found given by $r=R_+=1+(1-a^2)^{1/2}$, which exists
for $|a|<1$. Once the black hole
rotates ($a\neq0$), particles and photons are pushed to co-rotate with
the black hole due to frame-dragging effect. The co-rotation is obligatory
within the ergosphere. Let us note that the Kerr spin parameter is limited to an
equilibrium value by photon recapture from the disc ($a\simeq0.998$)
and by magnetic torques.\cite{Tho:1974,Kro:2005}

The existence of ISCO is a remarkable feature of motion in
strong gravity.\cite{Bar:1972} For a co-rotating equatorial disc,
\begin{equation}
R_{\rm ISCO} = 3+z_2-\big[\left(3-z_1)(3+z_1+2z_2\right)\big]^\frac{1}{2},
\label{velocity1}
\end{equation}
where $z_1 = 1+\alpha_+\alpha_-[\alpha_++\alpha_-]$,
$\alpha_{\pm}=(1{\pm}a)^\frac{1}{3}$, and $z_2 =
(3a^2+Z_1^2)^\frac{1}{2}$. The ISCO radius as function of spin spans the 
range of dimensionless radius
from $R_{\rm ISCO}=1$ (for $a=1$, i.e.\ a maximally co-rotating Kerr black hole) to $R_{\rm ISCO}=6$
(for $a=0$, a static case of Schwarzschild black hole) to $R_{\rm ISCO}=9$
($a=-1$, case of maximum counter-rotation). 
The velocity of prograde Keplerian circular motion is
\begin{equation}
 v^{(\phi)}=\Delta^{-1/2}\left(r^2-2ar^{1/2}+{a}^2\right)\left(r^{3/2}
  +{a}\right)^{-1}
\end{equation}
with respect to locally non-rotating observers (LNRF). The corresponding angular velocity
is $\Omega(r;a)=(r^{3/2}+{a})^{-1}$. Finally, to derive the interval measured
by a distant observer, one needs to consider the Lorentz factor
associated with the orbital motion,
\begin{equation}
 \Gamma = \frac{\left(r^{3/2}+{a}\right)\Delta^{1/2}}{r^{1/4}\;
 \sqrt{r^{3/2}-3r^{1/2}+2{a}}\;\sqrt{r^3+{a}^2r+2{a}^2}}\,.
\end{equation}

\begin{figure}[tbh!]
\begin{center}
\includegraphics[width=0.43\textwidth]{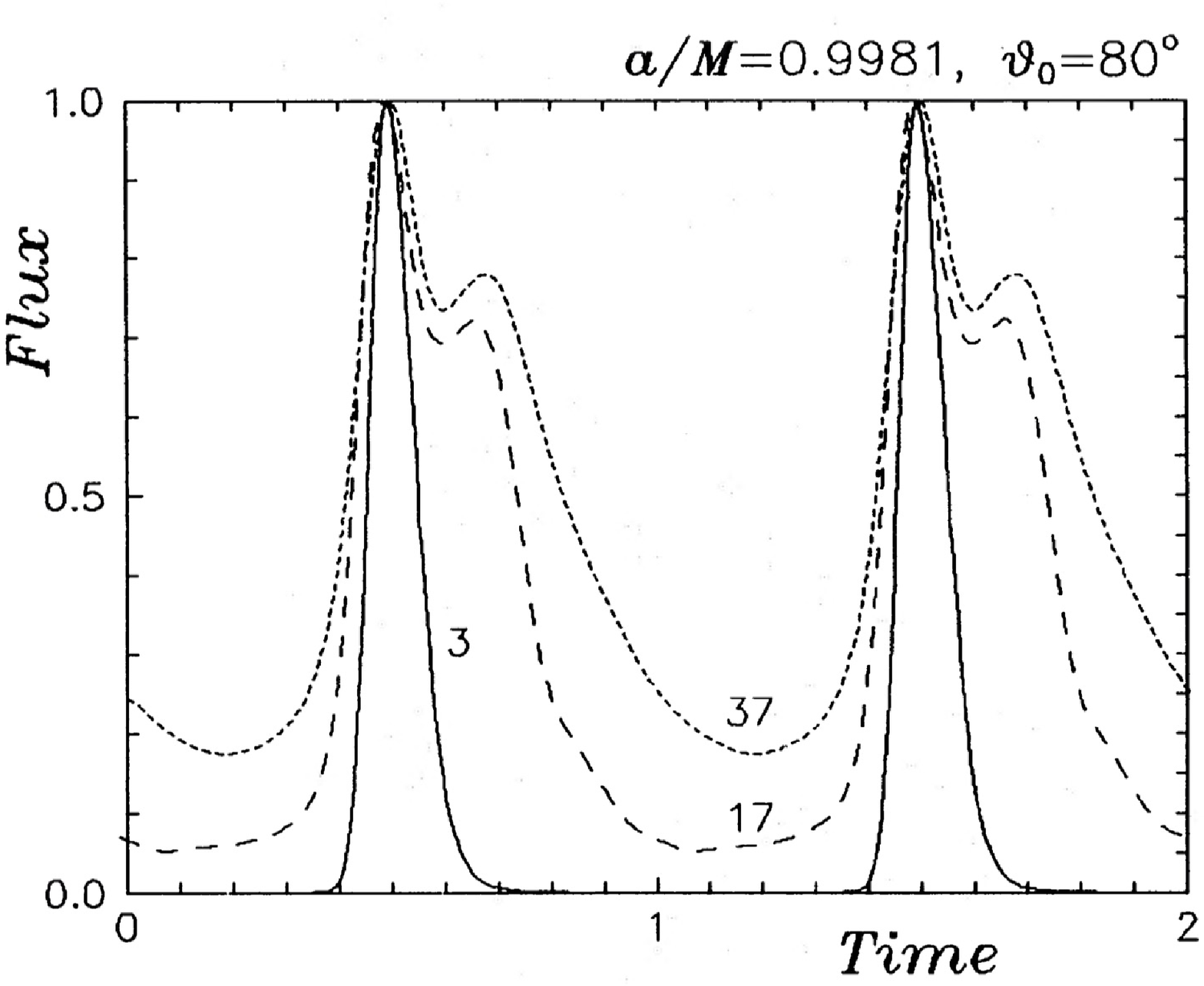}
\hfill
\includegraphics[width=0.43\textwidth]{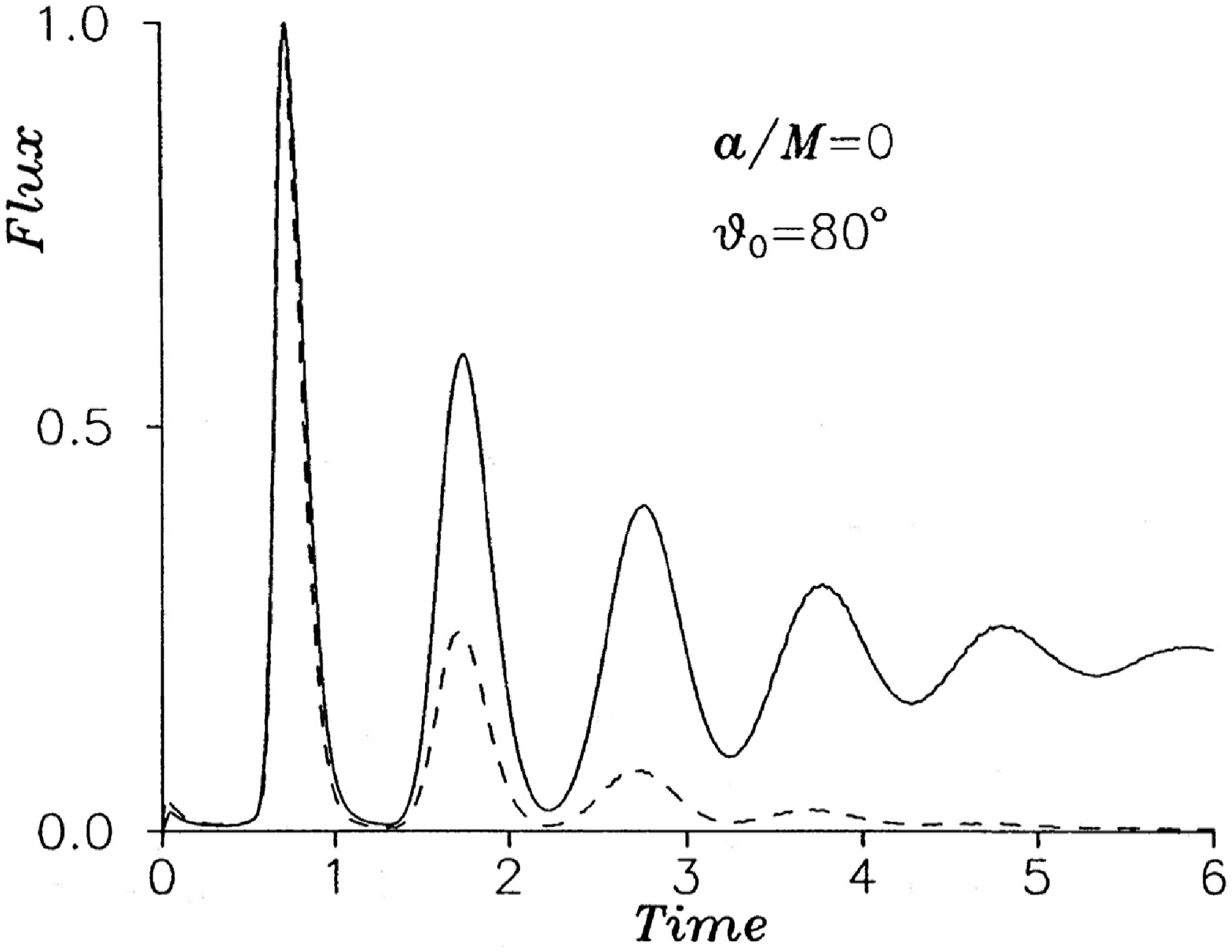}
\end{center}
\label{fig1}
\caption{Exemplary light curves of the signal from a spot orbiting on Keplerian
circular trajectory near Kerr black hole. Effects of General Relativity have been 
taken into account, which allows us to constrain the parameters of the system.
Left panel: radiation flux variation over a two revolutions. Parameters of the system 
are specified with the plot (radius of the orbits is given with each curve in units
of the black hole gravitational radius, $R_g=GM/c^2$). Right panel: Two cases
of tidally decaying spots due to differential rotation; (a) constant total emission from the
spot (solid line) vs.\ (b) exponentially decaying intrinsic emission with time (dashed line),
motivated by modelling the origin of the spot. Location of the spot is centered at
$r=6R_+$.\cite{Kar:1992}}
\end{figure}

Geodesic motion is determined by three constants of motion: the total
energy $\cal E$, the azimuthal component of angular momentum $L_{\rm
z}$, and Carter's constant $Q$. Null geodesics are
relevant to describe propagation of photons, and for these the number of free constants 
can be reduced:
$\xi = L_{z}/{\cal E}$, $\eta = Q/{\cal E}^{2}$. For photons
propagating from the accretion disk towards a distant observer, the
initial point is set at a given radius in the equatorial plane of the
black hole, whereas the final point is at radial infinity, along the
view angle of the observer. Relativistic effects become more prominent an high
inclination angles.\cite{Zak:2003,Kar:2010}

We employ geometrical optics and derive predicted light-curves from a source of light 
orbiting at a particular radius with the equatorial accretion disk. The equation for
photon rays (null geodesics) relates an emission point 
${\cal P}_0=(r,\theta,\phi)_{\rm em}$ near the black hole 
with the terminal point ($\alpha,\beta$) in observer's detector plane at spatial 
infinity.\cite{Car:1968,Kar:1992}
The rays can be integrated in terms of elliptic integrals,
\begin{equation}
\int_{R_{\rm em}}^R {\cal R}(r^\prime)^{-1/2}\,dr^\prime=
\int_{\theta_{\rm em}}^\theta \Theta(\theta^\prime)^{-1/2}\,d\theta=
\int_{\phi_{\rm em}}^\phi {\cal F}(\phi^\prime)^{-1}\,d\phi^\prime,
\end{equation}
where
\begin{eqnarray}
{\cal R}(r)&=&r^4+(a^2-\xi^2-\eta)r^2+2[\eta+(\xi-a)^2]r-a^2\eta,\\
\Theta(\theta)&=&\eta+a^2\cos^2\theta-\xi^2\cot^2\theta,\\
{\cal F}(\phi)&=&[2ar+(\Sigma-2r)\xi\csc^2\theta]\Delta^{-1},\\
\Delta(r)&=&r^2-2r+a^2,\quad \Sigma=r^2+a^2\cos^2\theta,\\
A(r,\theta)&=&(r^2+a^2)^2-\Delta(r)a^2\sin^2\theta.
\end{eqnarray}
The two constants of motion of the photon motion satisfy
\begin{eqnarray}
\xi&=&A^{1/2}\Sigma^{-1/2}E^{-1}\sin\theta\sin\alpha\sin\beta_{|{\cal P}_0}\,,\\
\eta&=&\Delta^{-1}(r^2+a^2-a\xi)^2-\Sigma E^{-2}\cos^2\alpha-\xi^2+2a\xi-{a^2}_{|{\cal P}_0}
\end{eqnarray}
with $EA^{1/2}=\Sigma^{1/2}\Delta^{1/2}+2ar\Sigma^{-1/2}\sin\theta\,\sin\alpha\,\sin\beta$.
Finally, time coordinate can be integrated in the form
\begin{equation}
t=\int_{t_0}^t \frac{A(r,\theta)-2ar\xi}{\Sigma(r,\theta)\,\Delta(r)}\;d\tau.
\end{equation}

To show exemplary solutions of the above-given equations and to illustrate the main effects of General Relativity, 
we plot several model light curves of an orbiting spot in Fig.\ \ref{fig1}. The main
parameters are the radial position of the center of the spot, view angle of a distant observer, and the Kerr parameter
of the black hole.\cite{Kar:1992,Dov:2004,Dov:2004b} One can distinguish two local maxima that dominate the light curve variation over the revolution
of the spot: (i) the Doppler boosting peak near the phase $\approx0.7$ (on the approaching side of the trajectory with respect to the observer,
with a certain time delay due to light-travel time), and (ii) the lensing peak near the phase $0.5$ (due to light bending
at the point of upper conjunction of the trajectory). Only one of the peaks may appear in some cases; the occurrence 
and the relative height of the two peaks depend on the model parameters. 
For spots on a close orbit around the black hole, the light rays are bent to the extent that the Doppler-peak follows immediately after 
the lensing peak, whereas for larger orbits the two peaks can be a quarter of an orbit apart from each other.
We propose that the changing
profile can be used to constrain the parameters of the system; in particular, this can allow us to determine the parameters
of the central black hole.\cite{Kar:2017}

\begin{figure}[tbh!]
\vspace*{-1em}
\begin{center}
\includegraphics[width=0.6\textwidth]{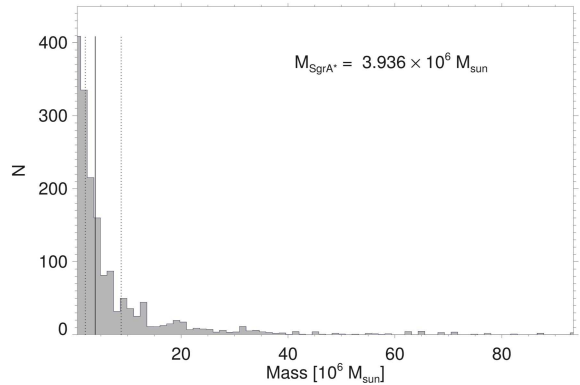}
\end{center}
\label{fig2}
\vspace*{-1em}
\caption{Histogram of the predicted mass of Sgr A* supermassive black hole derived from a combination of the four flares 
taken into account. The most likely mass predicted by the method is the median value indicated by the solid vertical line; figure
from Karssen et al.\ 2017.\cite{Kar:2017}}
\end{figure}

Solving the above-given equations usually requires to perform several steps numerically. Alternatively, let us mention that a number
of semi-analytical approximations have been formulated that allow us to examine selected aspects (e.g.\ extremal values of 
the redshift function from an orbiting spot).\cite{Pec:2005,Sem:2015,Def:2016} These can provide simple and practical estimates that are useful in analyzing the
redshifted narrow lines, however, the approximation approaches assume additional limitations that reduce the accuracy.

\section{The Method and Results}
Karssen et al. (2017) studied the light curves of the four bright flares of Sagittarius A*  (Sgr A*) in the center of the Milky Way. 
They allowed the mass of the black to vary as a free parameter and employed the model profiles to fit the most likely value. 
The flare time-scale in periods is matched with the actual duration of the observed flare in physical units, which provides a way 
to constrain the black hole mass. The quality and the volume of data for Sgr A* SMBH is the best, however, the same approach can be employed also
with other bright objects. In particular, the method has been tested with the Seyfert I galaxy RE J1034+396.\cite{Kar:2017}

The simulated light curves are normalized to the maximum flux of the light curve, which is set to be identical to the observed light curve.
To conduct a time efficient fit of the models to the data, we introduce a time shift, a flux density scaling factor and a flux density offset. 
For each light curve the ratio is considered between the number of data points of the light curve which belong to the flaring period (defined by
the requirement to on the flux to exceed 30 per cent of the maximum value) and the number of data points of the quiescent state. 
The main features of the flares are contained in the upper two-thirds of the flares. Every simulated light curve ratio is then compared 
to an observed one; quiescent state data points are removed from or added to the simulated data until the ratio is comparable. 

By the fitting process the theoretical light curves are lined up with the observed ones and the conversion factor is determined between 
geometrized time units of a particular model and the physical time-scale of the observation. This relation depends on the mass 
of the black hole. By gauging the intrinsic clock of the black hole in its gravitational units to the clocks of the observations in seconds, 
the mass of the black hole can be estimated. Each fit of a particular simulated light curve to a particular observed flare then results 
in an estimate for $M$. The simulated data is multiplied by a factor because in general the best-fit shape of the light curves does not 
depend on the initial normalization we inferred earlier. In Fig.\ \ref{fig2} we show the resulting mass predicted from a combination of 
the four brightest Sgr A* flares by; the y-axis of the diagram represents the number of model results that have predicted the particular mass
value. A peak in this diagram occurs where the most models predicting the corresponding value in the particular mass bin find
the minimum of best-fit statistics. Let us note that the resulting distributions are not normal distributions, hence we use the medians 
as a measure for the most probable mass, as indicated by the vertical solid line in the graph.

\section{Conclusions}
 The null geodesics were obtained by solving numerically the equations governing the photon propagation in Kerr space–time.
In this initial work we have not included additional complexities due to the intrinsic variation (decay) of the flares;
such a generalization is postponed for a future study, however, the necessary methodology has been already 
developed; cf.\ the right panel in Fig.\ \ref{fig1}, where an exponentially fading signal of a tidally sheared spot has
been studied.

We have outlined a method based on a comparison between the simulations with the bright X-ray flares from Sgr A*,
which gives an estimate on the mass of the black hole. Let us emphasize that the result does not depend on the uncertainties about 
the object distance, which otherwise complicates approaches to the mass estimation by different methods. 
On the other hand, the hereby described method has other underlying assumptions and it can be applied only 
if the light curves of the flare events are dominated by the effects of 
relativistic motion of the source orbiting the black hole. This may be adequate for a limited sub-sample of targets;
we have argued that the bright flares belong within the suitable category. For further details, see Karssen et al. (2017).\cite{Kar:2017}

\section*{Acknowledgments}
The authors acknowledge the Czech Science Foundation (GA\v{C}R) -- German Research Foundation (DFG) collaboration project No.\ 19-01137J 
and the Czech Ministry of Education, Youth and Sports collaboration program Kontakt ref.\ LTAUSA 17095.

\end{document}